
\magnification\magstep1

\openup .5\jot

\input mssymb
\def\hbar{\mathchar '26\mkern -9muh}

\catcode`@=11
\def\eqaltxt#1{\displ@y \tabskip 0pt
  \halign to\displaywidth {%
    \rlap{$##$}\tabskip\centering
    &\hfil$\@lign\displaystyle{##}$\tabskip\z@skip
    &$\@lign\displaystyle{{}##}$\hfil\tabskip\centering
    &\llap{$\@lign##$}\tabskip\z@skip\crcr
    #1\crcr}}
\def\eqallft#1{\displ@y \tabskip 0pt
  \halign to\displaywidth {%
    $\@lign\displaystyle {##}$\tabskip\z@skip
    &$\@lign\displaystyle{{}##}$\hfil\crcr
    #1\crcr}}
\catcode`@=12 

\def\half{{\textstyle {1 \over 2}}}

\def\pmb#1{\setbox0=\hbox{#1}  \kern-.025em\copy0\kern-\wd0
  \kern0.05em\copy0\kern-\wd0  \kern-.025em\raise.0433em\box0 }
\def\pmbh#1{\setbox0=\hbox{#1} \kern-.12em\copy0\kern-\wd0
            \kern.12em\copy0\kern-\wd0\box0}
\def\sqr#1#2{{\vcenter{\vbox{\hrule height.#2pt
      \hbox{\vrule width.#2pt height#1pt \kern#1pt
         \vrule width.#2pt}
      \hrule height.#2pt}}}}

\def\rchi{{\raise 2pt \hbox {$\chi$}}}
\def\rga{{\raise 2pt \hbox {$\gamma$}}}
\def\rg{{\raise 2 pt \hbox {$g$}}}

\def\({\left(}
\def\){\right)}
\def\<{\left\langle}
\def\>{\right\rangle}

\def\[{\left[}
\def\]{\right]}
\let\text=\hbox
\def\pt{\partial}

\def\lam{\lambda}

\def\sig{\sigma}

\hfuzz 6pt

\catcode`@=12 
\vskip .2 true in
\centerline {\bf Quantization of a Friedmann-Robertson-Walker model in N=1
Supergravity  }
\centerline {\bf with
Gauged Supermatter}
\vskip .1 true in
\centerline {A.D.Y. Cheng, P.D. D'Eath and
{\rm P.R.L.V. Moniz}\footnote*{{\rm e-mail address: prlvm10@amtp.cam.ac.uk}}}
\vskip .1 true in
\centerline {Department of Applied Mathematics and Theoretical Physics}
\centerline {University of Cambridge, Silver Street,Cambridge CB3 9EW, UK }
\vskip .2 true in
\centerline {\bf ABSTRACT}
\vskip .1 true in

{\sevenrm The theory of  N = 1
 supergravity with gauged supermatter is studied in the context of a  k = + 1
Friedmann
minisuperspace model.
It is found by imposing the Lorentz and supersymmetry constraints that there
are
{\seveni no} physical states in the particular $SU(2)$ model studied. }

\noindent

\magnification\magstep1

\openup .5\jot

\input mssymb
\def\hbar{\mathchar '26\mkern -9muh}

\catcode`@=11
\def\eqaltxt#1{\displ@y \tabskip 0pt
  \halign to\displaywidth {%
    \rlap{$##$}\tabskip\centering
    &\hfil$\@lign\displaystyle{##}$\tabskip\z@skip
    &$\@lign\displaystyle{{}##}$\hfil\tabskip\centering
    &\llap{$\@lign##$}\tabskip\z@skip\crcr
    #1\crcr}}
\def\eqallft#1{\displ@y \tabskip 0pt
  \halign to\displaywidth {%
    $\@lign\displaystyle {##}$\tabskip\z@skip
    &$\@lign\displaystyle{{}##}$\hfil\crcr
    #1\crcr}}
\catcode`@=12 

\def\half{{\textstyle {1 \over 2}}}

\def\pmb#1{\setbox0=\hbox{#1}  \kern-.025em\copy0\kern-\wd0
  \kern0.05em\copy0\kern-\wd0  \kern-.025em\raise.0433em\box0 }

\def\pmbh#1{\setbox0=\hbox{#1} \kern-.12em\copy0\kern-\wd0
            \kern.12em\copy0\kern-\wd0\box0}
\def\sqr#1#2{{\vcenter{\vbox{\hrule height.#2pt
      \hbox{\vrule width.#2pt height#1pt \kern#1pt
         \vrule width.#2pt}
      \hrule height.#2pt}}}}

\def\rchi{{\raise 2pt \hbox {$\chi$}}}
\def\rga{{\raise 2pt \hbox {$\gamma$}}}
\def\rrho{{\raise 2pt \hbox {$\rho$}}}

\def\({\left(}
\def\){\right)}
\def\<{\left\langle}
\def\>{\right\rangle}

\def\[{\left[}
\def\]{\right]}
\let\text=\hbox
\def\pt{\partial}

\def\lam{\lambda}

\def\sig{\sigma}


The subjects of   supersymmetric quantum gravity and cosmology have achieved a
considerable number
of very interesting results and conclusions during the last ten years or so
[2,3].
 Our objective here is to  study a  $k=1$  supersymmetric FRW mini-superspace
quantum
 cosmological model  with a family of spin-0 as well as spin-1 gauge
fields together with their odd (anti-commuting) spin-$\half$ partners with zero
analytic potential $ P \( \Phi^I \) $.
 The supersymmetry constraints will be  derived from the reduced theory with
supermatter.
 Subsequently,  we solve for the components of the wave function using the
 quantum constraints. We will then find that there are {\it no} solutions for
the
 quantum states of the FRW universe analysed here.

Let us begin by specifying our model in some detail.
The Lagrangian of the theory studied here is given in Eq. (25.12) of ref. [1];
 it is too long to write out here. We choose the geometry
to be that of a $ k = + 1 $ Friedmann model with $ S^3 $ spatial sections,
which are the spatial orbits of $G=SO(4)$ -- the group of homogeneity and
isotropy. The tetrad of the four-dimensional theory can be  taken to be:
$$ e_{a\mu} = \pmatrix {
N (\tau) &0 \cr
0  &a E_{\hat a i} \cr }~,
\ \ \ e^{a \mu} = \pmatrix {
N (\tau)^{-1} & 0 \cr
0 &a (\tau)^{-1} E^{\hat a i} \cr }$$
where $ \hat a $ and $ i $ run from 1 to 3.
$ E_{\hat a i} $ is a basis of left-invariant 1-forms on the unit $ S^3 $
with volume $ \sig^2 = 2 \pi^2 $. The spatial tetrad $ e^{AA'}_{~~~~i} $
satisfies the relation
$$ \pt_i e^{AA'}_{~~~~j} - \pt_j e^{AA'}_{~~~~i} = 2 a^2 e_{ijk} e^{AA'k} $$
as a consequence of the group structure of SO(3), the isotropy
(sub)group.

This Ansatz reduces the number of degrees of freedom provided by $ e_{AA'
\mu} $. If supersymmetry invariance is to be retained, then we need an
Ansatz for $ \psi^A_{~~\mu} $ and $ \bar\psi^{A'}_{~~\mu} $ which reduces the
number of fermionic degrees of freedom, so that there is equality between the
number of bosonic and fermionic degrees of freedom. One is naturally led to
take $
\psi^A_{~~0} $ and $ \bar\psi^{A'}_{~~0} $ to be functions of time only. We
further take
$$
\psi^A_{~~i} = e^{AA'}_{~~~~i} \bar\psi_{A'}~, ~
\bar\psi^{A'}_{~~i} = e^{AA'}_{~~~~i} \psi_A~  $$
where we introduce the new spinors $ \psi_A $ and $ \bar\psi_{A'} $ which
are functions of time only. [It is possible to justify the  above Ansatz by
requiring that the form of the tetrad be preserved under suitable
homogeneous supersymmetry transformations. [2]

Now, consider the supermatter fields.
The scalar super-multiplet, consisting of a complex massive scalar
field $ \phi $ and massive spin-$\half$ field $ \rchi, \bar\rchi$
are chosen to be spatially homogeneous, depending only on time. The
odd spin-$\half $ partner $ \( \lam^{(a)}, \bar \lam^{(a)}\) $, $a=1,2,3$, is
chosen to
depend only on time as well. As far as the vector field $ A^{(a)}_\mu $
is concerned we adopt here the Ansatz formulated in ref. [4] and choose
$$
{\bf A}_{\mu}(t)~\omega^{\mu}  =
{{f(t)}\over {2}}
{\cal T}_{c}\omega^c $$
where $\{\omega^{\mu}\}$ represents the moving coframe
$\{\omega^{\mu}\} = \{ dt,\omega^b\}~,~(b=1,2,3)~$,  of
one-forms, invariant under the left action of $SU(2)$ and ${\cal T}_{a}$
are the generators of the $SU(2)$ gauge group.
Notice in the above form for the gauge field  the $A_0$ component is taken
to be identically zero. Thus, we will have in our FRW case a gauge  constraint
$Q^{(a)}=0$.

Using the Ans\"atze previously described,
the action of the full theory (Eq. (25.12) in ref. [1]) can be reduced to one
with a
 finite number of degrees of freedom. Notice that  with our choice of gauge
group $SU(2)$ and compact K\"ahler manifold, itdirectly follows that the
analytical potential $P(\Phi^{I})$ is zero [5]

Let us here solve explicitly the corresponding quantum
supersymmetry constraints.
 First we need to redefine the  fermionic fields, $ \rchi_{A} $,  $ \psi_{A} $
and $\lambda_{A}$
in order to simplify the Dirac
brackets , following  the steps described in ref. [6]:
$$ \hat \rchi_{A} = {\sigma a^{3 \over 2} \over 2^{1 \over 4} (1 + \phi \bar
\phi)} \rchi_{A}~, ~
\hat{ \bar \rchi}_{A'} = {\sigma a^{3 \over 2} \over 2^{1 \over 4} (1 + \phi
\bar \phi)} \bar \rchi_{A'}~,~\pi_{\hat \rchi_{A}} = -i n_{AA'} \hat{ \bar
\rchi}^{A'}~, ~
 \pi_{\hat{ \bar \rchi}_{A'}} = -i n_{AA'} \hat \rchi^{A} $$
$$ \hat \psi_{A} = {\sqrt{3} \over 2^{1 \over 4}} \sigma a^{3 \over 2}
\psi_{A}~, ~
\hat{\bar \psi}_{A'} = {\sqrt{3} \over 2^{1 \over 4}} \sigma a^{3 \over 2} \bar
\psi_{A'} ~,
\pi_{\hat{ \psi}_{A}} = in_{AA'} \hat{\bar \psi}^{A'} ~,~
\pi_{\hat{\bar  \psi}_{A'}} = in_{AA'} \hat \psi^{A}~ $$
$$ \hat \lambda^{(a)}_{~A} = {\sigma a^{ 3 \over 2} \over 2^{ 1 \over 4}
}\lambda^{(a)}_{~A}, ~~
\hat{\bar \lambda}^{(a)}_{~A'} = {\sigma a^{ 3 \over 2} \over 2^{ 1 \over 4}
}\bar \lambda^{(a)}_{~A'}
{}~, \pi_{\hat{ \lambda}^{(a)}_{~A}} = -in_{AA'} \hat{\bar \lambda }^{(a)A'}
{}~,~
\pi_{\hat{\bar \lambda}^{(a)}_{~A'}} = -in_{AA'} \hat \lambda^{(a)A} $$
The Dirac brackets are:
$$ [ \hat \rchi_{A} , \hat{\bar \rchi}_{A'} ]_{*} = -i n_{AA'}~, [\hat \psi_{A}
, \hat{\bar \psi}_{A'}]_{*} = in_{AA'}~,  [\hat \lambda^{(a)}_{~A}, \hat{\bar
\lambda}^{(a)}_{~A'}]_{*} = - i \delta^{ab} n_{AA'} $$
$$ [a , \pi_{a}]_{*} = 1~, ~ [\phi, \pi_{\phi}]_{*} = 1~,
 ~[\bar \phi, \pi_{\bar \phi}]_{*} = 1~, ~[f, \pi_{f}] = 1 $$
and the rest of the brackets are zero.
It is simpler to describe the theory using only (say) unprimed spinors, and, to
this end, we define
$$ \bar \psi_{A} = 2 n_{A}^{~B'} \bar \psi_{B'}~, ~
 \bar \rchi_{A} = 2 n_{A}^{~B'} \bar \rchi_{B'} ~, ~\bar \lambda^{(a)}_{~A} = 2
n_{A}^{~B'}
 \bar \lambda^{(a)}_{~B'}  $$
with which the new Dirac brackets are
$$ [\rchi_{A}, \bar \rchi_{B}]_{*} = -i \epsilon_{AB}~, ~
 [\psi_{A}, \bar \psi_{B}]_{*} = i \epsilon_{AB} ~, [\lambda^{(a)}_{~A}, \bar
\lambda^{(a)}_{~A'}]_{*} =
 - i \delta^{ab} \epsilon _{AB}  $$
The rest of the brackets remain unchanged. Hence the only non-zero
(anti-)commutator relations are:
$$ \{\lambda^{(a)}_{~A}, \lambda^{(b)}_{~B} \} = \delta^{ab} \epsilon_{AB} ~,~
\{\rchi_{A} , \bar \rchi_{B} \} = \epsilon_{AB}~, ~
\{\psi_{A} , \bar \psi_{B}\}  = - \epsilon_{AB} $$
$$ [a , \pi_{a}] = [\phi , \pi_{\phi}] = [\bar \phi, \pi_{\bar \phi}] = [f,
\pi_{f}] = i  $$
Here we choose $ (\rchi_{A} , \psi_{A} , a , \phi , \bar \phi) $ to be the
coordinates of the configuration space, and
$ \bar \rchi_{A} , \bar \psi_{A} , \pi_{a} , \pi_{\phi} , \pi_{\bar \phi} $ to
be the momentum operators in this representation.
Hence

$$  \lambda^{a}_{~A} \rightarrow -{\pt \over \pt \bar \lambda^{(a)A}} ~,~
 \bar \rchi^{A} \rightarrow -{\pt \over \pt \rchi^{A}} ,~ \bar \psi_{A}
\rightarrow {\pt \over \pt \psi^{A}} $$
 $$ \pi_{a} \rightarrow {\pt \over \pt a} , ~\pi_{\phi} \rightarrow -i {\pt
\over \pt \phi},~
 \pi_{\bar \phi} \rightarrow -i {\pt \over \pt \phi} ~,~ \pi_{f} \rightarrow -i
{\pt \over \pt f}  $$
Following the ordering used in ref. [2],  we put all the fermionic derivatives
in  $S_{A}$ on the right.
 In $ \bar S_{A} $, all the fermonic
derivatives are on the left. Implementing all these redefinitions, the
 supersymmetry constraints  have the differential operator form

$$ S_{A} = -{i \over \sqrt{2}} (1 + \phi \bar \phi) \rchi_{A} {\pt \over  \pt
\phi}
 - {1 \over 2 \sqrt{6}} a \psi_{A} {\pt \over \pt a} - \sqrt{3 \over 2}
\sigma^{2}a^{2} \psi_{A}
- {5i \over 4 \sqrt{2}} \bar \phi \rchi_{A} \rchi^{B} {\pt \over \pt
\rchi^{B}}$$
$$-{1 \over 8 \sqrt{6}} \psi_{B} \psi^{B} {\pt \over \pt \psi^{A}}
- {i \over 4 \sqrt{2}} \bar \phi \rchi_{A} \psi^{B} {\pt \over \pt \psi^{B}}
-{5 \over 4 \sqrt{6}} \rchi_{A} \psi^{B}{\pt \over \pt \rchi^{B}} + {\sqrt{3}
\over 4 \sqrt{2}} \rchi^{B} \psi_{B} {\pt \over \pt \rchi^{A}} $$
$$+ {1 \over 2 \sqrt{6}} \psi_{A} \rchi^{B} {\pt \over \pt \rchi^{B}} + {1
\over 3 \sqrt{6}} \sigma^{a}_{~AB'} \sigma^{bCC'} n_{D}^{~B'} n^{B}_{~C'} \bar
\lambda^{(a)D} \psi_{C} {\pt \over \pt
 \bar \lambda^{(b)B}}$$
 $$ + {1 \over 6 \sqrt{6}} \sigma^{a}_{~AB'} \sigma^{bBA'} n_{D}^{~B'}
n^{E}_{~A'} \bar \lambda^{(a)D} \bar \lambda^{(b)}_{~B} {\pt \over \pt
\psi^{E}} -{ 1 \over 2 \sqrt{6}} \psi_{A} \bar \lambda^{(a)C} {\pt \over \bar
\lambda^{(a)C}} + {3 \over 8 \sqrt{6}} \bar \lambda^{a}_{~A} \lambda^{(a)C}
{\pt \over \pt \psi^{C}} $$
$$ + {1 \over 2 \sqrt{2}} \sigma^{2} a^{3} g \bar D^{(a)} \bar \lambda^{a}_{~A}
- {1 \over 4 \sqrt{6}} \psi^{C} \bar \lambda^{(a)}_{~C} {\pt \over \pt \bar
\lambda^{(a)A}} + {\sigma^{2} a^{2} g f \over \sqrt{2} (1 + \phi \bar \phi)}
\sigma^{a}_{~AA'} n^{BA'} \bar X^{(a)} \chi_{B}$$
$$ + \sigma^{a}_{~AA'} n^{BA'}  \bar \lambda^{(a)}_{~B} \left( -{ \sqrt{2}
\over 3}  { \pt \over \pt f} + {1 \over 8 \sqrt{2}} (1 -(f-1)^{2}) \sigma^{2}
a^{4} \right)  $$

We now proceed to find the wavefunction of our model.
The Lorentz constraint $ J_{AB} $ is easy to solve. It tells us that the
wave function should be a  Lorentz scalar.  We can  see that the most general
form of the wave function which satisfies the Lorentz constraint is

$$ \Psi = A + iB \psi^{C} \psi_{C} + C \psi^{C} \rchi_{C} + iD \rchi^{C}
\rchi_{C} + E \psi^{C} \psi_{C} \rchi^{C} \rchi_{C} $$
$$+ c_{a} \bar \lambda^{(a)C} \rchi_{C} + d_{a}  \bar \lambda^{(a)C} \psi_{C} +
c_{ab} \bar \lambda^{(a)C}  \bar \lambda^{(b)}_{~C}
+ e_{a}  \bar \lambda^{(a)C} \rchi_{C} \psi^{D} \psi_{D} $$
$$+ f_{a} \bar \lambda^{(a)C} \psi_{C}  \rchi^{D} \rchi_{D} + d_{ab}  \bar
\lambda^{(a)C} \rchi_{C} \bar \lambda^{(a)D} \rchi_{D}
+ e_{ab}  \bar \lambda^{(a)C}  \bar \lambda^{(b)}_{~C} \psi^{D} \psi_{D} $$
$$+ f_{ab} \bar \lambda^{(a)C}  \bar \lambda^{(b)}_{~C} \rchi^{D} \rchi_{D}
+g_{ab} \bar \lambda^{(a)C}  \bar \lambda^{(b)}_{~C} \rchi^{D} \psi_{D}
 + c_{abc}  \bar \lambda^{(a)C}  \bar \lambda^{(b)}_{~C} \bar \lambda^{(c)D}
\psi_{D}  $$
$$+ d_{abc} \bar \lambda^{(a)C}  \bar \lambda^{(b)}_{~C} \bar \lambda^{(c)D}
\rchi_{D} + c_{abcd} \bar \lambda^{(a)C}  \bar \lambda^{(b)}_{~C}
\bar \lambda^{(c)D}  \bar \lambda^{(d)}_{~D} + h_{ab}  \bar \lambda^{(a)C}
\bar \lambda^{(b)}_{~C} \psi^{D} \psi_{D} \rchi^{E} \rchi_{E} $$
$$+ e_{abc} \bar \lambda^{(a)C}  \bar \lambda^{(b)}_{~C}  \bar \lambda^{(c)D}
\rchi_{D} \psi^{E} \psi_{E} + f_{abc} \bar \lambda^{(a)C}  \bar
\lambda^{(b)}_{~C} \bar \lambda^{(c)D} \psi_{D} \rchi^{E} \rchi_{E} + d_{abcd}
\bar \lambda^{(a)C}  \bar \lambda^{(b)}_{~C} \bar \lambda^{(c)D}  \bar
\lambda^{(d)}_{~D}  \psi^{E} \psi_{E}$$
$$ + e_{abcd} \bar \lambda^{(a)C}  \bar \lambda^{(b)}_{~C} \bar \lambda^{(c)D}
\bar \lambda^{(d)}_{~D} \rchi^{E} \rchi_{E} + f_{abcd} \bar \lambda^{(a)C}
\bar \lambda^{(b)}_{~C} \bar \lambda^{(c)D}  \bar \lambda^{(d)}_{~D} \psi^{E}
\rchi_{E}  + g_{abcd} \bar \lambda^{(a)C}  \bar \lambda^{(b)}_{~C} \bar
\lambda^{(c)D} \psi_{D} \bar \lambda^{(d)E} \rchi_{E} $$
$$ \mu_{1}  \bar \lambda^{(2)C}  \bar \lambda^{(2)}_{~C} \bar \lambda^{(3)D}
\bar \lambda^{(3)}_{~D} \bar \lambda^{(1)E} \psi_{E}
+ \mu_{2}  \bar \lambda^{(1)C}  \bar \lambda^{(1)}_{~C} \bar \lambda^{(3)D}
\bar \lambda^{(3)}_{~D} \bar \lambda^{(2)E} \psi_{E}
+ \mu_{3}  \bar \lambda^{(1)C}  \bar \lambda^{(1)}_{~C} \bar \lambda^{(2)D}
\bar \lambda^{(2)}_{~D} \bar \lambda^{(3)E} \psi_{E}  $$
$$+ \nu_{1} \bar \lambda^{(2)C}  \bar \lambda^{(2)}_{~C} \bar \lambda^{(3)D}
\bar \lambda^{(3)}_{~D} \bar \lambda^{(1)E} \rchi_{E}
+ \nu_{2}  \bar \lambda^{(1)C}  \bar \lambda^{(1)}_{~C} \bar \lambda^{(3)D}
\bar \lambda^{(3)}_{~D} \bar \lambda^{(2)E} \rchi_{E}
+ \nu_{3} \bar \lambda^{(1)C}  \bar \lambda^{(1)}_{~C} \bar \lambda^{(2)D} \bar
\lambda^{(2)}_{~D} \bar \lambda^{(3)E} \rchi_{E} $$
$$+ F \bar \lambda^{(1)C}  \bar \lambda^{(1)}_{~C} \bar \lambda^{(2)D} \bar
\lambda^{(2)}_{~D} \bar \lambda^{(3)E} \bar \lambda^{(3)}_{~E}
+ h_{abcd} \bar \lambda^{(a)C}  \bar \lambda^{(b)}_{~C} \bar \lambda^{(c)D}
\bar \lambda^{(d)}_{~D} \psi^{E} \psi_{E} \rchi^{F} \rchi_{F} $$
$$+ \delta_{1} \bar \lambda^{(2)C}  \bar \lambda^{(2)}_{~C} \bar \lambda^{(3)D}
\bar \lambda^{(3)}_{~D} \bar \lambda^{(1)E} \psi_{E}
\rchi^{F} \rchi_{F} + \delta_{2} \bar \lambda^{(1)C}  \bar \lambda^{(1)}_{~C}
\bar \lambda^{(3)D} \bar \lambda^{(3)}_{~D} \bar \lambda^{(2)E} \psi_{E}
\rchi^{F} \rchi_{F} $$
$$ + \delta_{3} \bar \lambda^{(1)C}  \bar \lambda^{(1)}_{~C} \bar
\lambda^{(2)D} \bar \lambda^{(2)}_{~D} \bar \lambda^{(3)E} \psi_{E}
\rchi^{F} \rchi_{F} + \gamma_{1} \bar \lambda^{(2)C}  \bar \lambda^{(2)}_{~C}
\bar \lambda^{(3)D} \bar \lambda^{(3)}_{~D} \bar \lambda^{(1)E} \rchi_{E}
\psi^{F} \psi_{F} $$
$$ + \gamma_{2}  \bar \lambda^{(1)C}  \bar \lambda^{(1)}_{~C} \bar
\lambda^{(3)D} \bar \lambda^{(3)}_{~D} \bar \lambda^{(2)E} \rchi_{E}
\psi^{F} \psi_{F}  + \gamma_{3} \bar \lambda^{(1)C}  \bar \lambda^{(1)}_{~C}
\bar \lambda^{(2)D} \bar \lambda^{(2)}_{~D} \bar \lambda^{(3)E} \rchi_{E}
\psi^{F} \psi_{F} $$
$$ + G  \bar \lambda^{(1)C}  \bar \lambda^{(1)}_{~C} \bar \lambda^{(2)D} \bar
\lambda^{(2)}_{~D} \bar \lambda^{(3)E} \bar \lambda^{(3)}_{~E} \psi^{F}
\psi_{F} + H  \bar \lambda^{(1)C}  \bar \lambda^{(1)}_{~C} \bar \lambda^{(2)D}
\bar \lambda^{(2)}_{~D} \bar \lambda^{(3)E} \bar \lambda^{(3)}_{~E}
\rchi^{F} \rchi_{F} $$
$$ + I  \bar \lambda^{(1)C}  \bar \lambda^{(1)}_{~C} \bar \lambda^{(2)D} \bar
\lambda^{(2)}_{~D} \bar \lambda^{(3)E} \bar \lambda^{(3)}_{~E}
\rchi^{F} \psi_{F} + K  \bar \lambda^{(1)C}  \bar \lambda^{(1)}_{~C} \bar
\lambda^{(2)D} \bar \lambda^{(2)}_{~D} \bar \lambda^{(3)E} \bar
\lambda^{(3)}_{~E}
\psi^{F} \psi_{F} \rchi^{G} \rchi_{G} $$
where $A$, $B$, $C$, $D$, $E$ etc  are functions of $a$, $\phi$ and $\bar \phi$
only.
This Ansatz contains all allowed combinations of the fermionic fields and
 is the most general Lorentz invariant function.

The next step is to solve the supersymmetry constraints $ S_{A} \Psi = 0 $ and
$ \bar S_{A'} \Psi = 0 $. Since each order in  fermionic variables is
independent,
the number of constraint equations will be very high. Their full
analysis is quite tedious and to write all the terms would overburden the
reader.
Let us show some examples of the calculations  involved in
solving the $S_{A} \Psi = 0$ constraint.

Consider the terms linear in $\rchi_{A}$:

$$ \left[ -{i \over \sqrt{2}} (1 + \phi \bar \phi) { \pt A \over \pt \phi}
\right] \rchi_{A} +
{ \sigma^{2} a^{2} g f \over \sqrt{2} (1 + \phi \bar \phi) } \sigma^{a}_{~AA'}
n^{BA'} \bar X^{(a)} A \rchi_{B} = 0. $$
Since this is true for all $\rchi_{A}$, the above equation becomes
$$ \left[ -{i \over \sqrt{2}} (1 + \phi \bar \phi) { \pt A \over \pt \phi}
\right] \epsilon_{A}^{~B} +
{ \sigma^{2} a^{2} g f \over \sqrt{2} (1 + \phi \bar \phi) } \sigma^{a}_{~AA'}
n^{BA'} \bar X^{(a)} A  = 0. $$
Mutliplying the whole equation by $n_{BB'}$ and using the relation $n_{BB'}
n^{BA'} = {1 \over 2} \epsilon_{B'}^{~A'}$,
we can see that the two terms are independent of each other since the $\sigma$
matrices are orthogonal to the n matrix.
Thus, we conclude that $ A=0. $
As we proceed, this pattern keeps repeating itself. Some equations
show that the coefficients have some symmetry
properties. For example, $d_{ab} = 2 g_{ab}$. But when these two terms
 are combined with each other, they become zero. It can be
seen as follows,

$$ d_{ab}  \bar \lambda^{(a)C} \rchi_{C} \bar \lambda^{(a)D} \psi_{D} +
g_{ab} \bar \lambda^{(a)C}  \bar \lambda^{(b)}_{~C} \rchi^{D} \psi_{D} $$
$$ = 2 g_{ab} \bar \lambda^{(a)C} \bar \lambda^{(b)D} \rchi_{D} \psi_{C} +
g_{ab} \bar \lambda^{(a)C}  \bar \lambda^{(b)}_{~C} \rchi^{D} \psi_{D} $$
$$ = - g_{ab} \bar \lambda^{(a)C}  \bar \lambda^{(b)}_{~C} \rchi^{D} \psi_{D} +
g_{ab} \bar \lambda^{(a)C}  \bar \lambda^{(b)}_{~C} \rchi^{D} \psi_{D} $$
using the property that $g_{ab} = g_{ba}$ and the spinor identity $ \theta_{AB}
= {1 \over 2} \theta_{C}^{~C}
\epsilon_{AB} $ where $\theta_{AB} $ is anti-symetric in the two indices.
 The same property applies to the terms with coefficients $ f_{abcd}$ and
$g_{abcd}$.
 Other equations imply that the
coefficients $c_{abc} ~,~ d_{abc} ~,~ c_{abcd} ~,~ e_{abc} ~,~ f_{abc} ~,~
d_{abcd} ~$,
$~ e_{abcd} ~,$ $h_{abcd}$ are totally
symmetric in their indices. This then leads to the terms cancelling with each
other,
 as can easily be shown. In the end, considering both the
$S_{A} \Psi = 0$ and $\bar S_{A} \Psi = 0$  constraints, we are left with the
surprising
result that the wave function must be zero in order to satisfy
the quantum constraints.

\medbreak

To summarise, we  have applied the canonical formulation
of  the more general theory of $ N = 1 $
 supergravity with supermatter [6] to a $k=+1$  FRW mini-superspace model,
subject to suitable
 Ans\"atze for the the gravitational field,
 gravitino field and the gauge vector field $A^{a}_{\mu}$ as well as the scalar
 fields and corresponding fermionic partners.
After a dimensional reduction, we derived the supersymmetric constraints
for our one-dimensional model.
We then solved the Lorentz and supersymmetry constraints for the case of a
two-dimensional spherically symmetric  K\"ahler manifold.
 We found that there are {\it no} physical states in this model.
 A similar conclusion was also obtained in ref. [7,8]
where no matter but a cosmological constant term was present.
All this seems to suggest that as one introduces more terms in a locally
supersymmetric
 action,giving more field modes with associated mixing, then  the
 constraints impose severe restrictions on the possible allowed states,
assuming homogeneity
and isotropy. This is not to say that there might not be many inhomogenous
states.

\noindent
{\bf ACKNOWLEDGEMENTS}

The authors are grateful  to S.W. Hawking for helpful
comments and to J. Mour\~ao for useful conversations.
Discussions with R. Capovilla and O. \'Obregon are also acknowledged.
A.D.Y.C. thanks the Croucher Foundation of Hong Kong for financial support.
P.R.L.V.M.  gratefully acknowledges the support of
a Human Capital and Mobility
Fellowship from the European Union (Contract ERBCHBICT930781).

\vskip .2 true in
\noindent
{\bf REFERENCES}

{\rm

\advance\leftskip by 4em
\parindent = -4em

[1]  J. Wess and J. Bagger, {\it Supersymmetry and Supergravity},
2nd.~ed. (Princeton University Press, 1992).

[2] P.D. D'Eath and D.I. Hughes, Nucl.~Phys.~B {\bf 378}, 381 (1992).

[3] J.Bene and R. Graham, gr-qc/9306017

[4] P.V. Moniz and J. Mour\~ao, Class.  Quantum Grav. {\bf 8}, (1991) 1815 and
references therein.

[5] E. Witten and J. Bagger, Phys.~Lett.~B {\bf 115} (1982).

[6] A.D.Y. Cheng, P.D. D'Eath and P.R.L.V. Moniz, DAMTP R94/13

[7] R. Capovilla and O. Obregon, Phys. Rev. D{\bf 49} (1994) 6562.

[8]  A.D.Y. Cheng, P.D. D'Eath and
P.R.L.V. Moniz, Phys. Rev. D{\bf 49} (1994) 5246.

\ \ \ }

\bye